\documentclass[twocolumn,aps]{revtex4}
\usepackage{amsmath}
\usepackage{amssymb}
\usepackage{comment}

\newcommand {\bea}{\begin{eqnarray}}
\newcommand {\eea}{\end{eqnarray}}
\newcommand {\be}{\begin{equation}}
\newcommand {\ee}{\end{equation}}

\usepackage{graphicx}

\begin{document}

%\draft
%\preprint{xx}

\title{Neutrino Emission from Ungapped Quark Matter}

\author{Thomas~Sch\"afer$^{1,2}$ and Kai~Schwenzer$^1$}

\affiliation{
$^1$Department of Physics, North Carolina State University,
Raleigh, NC 27695\\
%$^2$Department of Physics, SUNY Stony Brook, 
%Stony Brook, NY 11794\\ 
$^2$Riken-BNL Research Center, Brookhaven National 
Laboratory, Upton, NY 11973}

\begin{abstract}
We study neutrino emission from a normal, ungapped, quark 
phase in the core of a compact star. Neutrino emission 
from non-interacting quark matter leads to an emissivity 
that scales as $\epsilon\sim T^7$. We show that the emissivity 
is enhanced by a combination of Fermi liquid and non-Fermi 
liquid effects. Fermi liquid effects lead to an emissivity 
that scales as $\epsilon\sim \alpha_s T^6$, as originally 
shown by Iwamoto. We demonstrate that non-Fermi liquid effects 
further enhance the rate, leading to $\epsilon\sim \alpha_s^3 
T^6\log(m/T)^2$, where $m$ is the electric screening scale 
and $m\gg T$ under the conditions found in compact stars. We show, 
however, that combined with non-Fermi liquid effects in the specific
heat the enhancement in the emissivity only leads to a modest
reduction in the temperature of the star at late times. Our 
results confirm existing bounds on the presence of ungapped
quark matter in compact stars. We also discuss neutrino 
emission from superconducting phases with ungapped fermionic
excitations. 
\end{abstract}
\maketitle

\newpage

%%%%%%%%%%%%%%%%%%%%%%%%%%%%%%%%%%%%%%%%%%%%%%%%%%%%%%%%%%%%%%%%%%%%%%%%%
\section{Introduction}
\label{sec_intro}
%%%%%%%%%%%%%%%%%%%%%%%%%%%%%%%%%%%%%%%%%%%%%%%%%%%%%%%%%%%%%%%%%%%%%%%%%

Compact stars provide a unique opportunity to study cold and dense 
strongly interacting matter and its rich phase structure. An important
source of information about the structure of compact stars is their 
cooling behavior \cite{Pethick:1991mk,Lattimer:2004pg,Yakovlev:2004iq,Weber:2004kj,Page:2004fy}. 
For the first $\sim 10^5$ years after the star is born neutrino emission 
from the bulk is the most efficient energy loss mechanism. Since the 
matter in the interior of the star is almost degenerate the exact 
nature of the cooling mechanism is very sensitive to the structure 
of the low energy excitations. As a result the cooling behavior 
places important constraints on the phase diagram of dense matter. 

 Cooling mechanisms are generally grouped into fast processes 
with an emissivity $\epsilon\sim T^6$ and slow processes with 
$\epsilon\sim T^8$. Slow processes include the modified Urca
process \cite{Friman:zq} and neutrino bremsstrahlung. Examples
for fast mechanisms are the direct Urca process \cite{Lattimer:ib}, 
neutrino emission from pion or kaon condensates 
\cite{Maxwell:1977,Thorsson:1995rk}, and neutrino emission 
from an ungapped quark phase \cite{Iwamoto:eb}. In a fully gapped 
phase of nuclear or quark matter the emissivity is exponentially 
small. Near the critical temperature for superconductivity the 
emissivity is dominated by pair breaking and recombination with a 
temperature behavior that is intermediate between fast and slow 
mechanisms \cite{Flowers:1976,Jaikumar:2001hq}.

 In this paper we wish to focus on the emissivity of a possible
quark matter phase. At asymptotically high baryon density quark
matter is in the color-superconducting CFL phase \cite{Alford:1998mk}.
In this phase all quark excitations are gapped and the emissivity 
is dominated by exponentially small processes involving massive 
pseudo-Goldstone modes \cite{Jaikumar:2002vg,Reddy:2002xc,Kundu:2004mz}. 
At densities that are relevant to compact stars distortions of the 
CFL due to the non-zero strange quark mass cannot be neglected 
and the phase structure is much more complicated. Most of the 
phases that have been proposed in the literature involve quarks
with vanishing or very small gaps. Examples are the 2SC phase 
\cite{Bailin:1984bm,Alford:1998zt,Rapp:1998zu} in which one 
of the colors remains ungapped, single flavor spin-one superconductors
that have very small gaps \cite{Schafer:2000tw,Schmitt:2003aa}, and
gapless CFL or 2SC phases \cite{Alford:2003fq,Shovkovy:2003uu}.

  In the following we will concentrate on the neutrino emissivity 
of ungapped quark matter. This problem is important both in 
connection with the (almost) gapless phases mentioned above and
as a benchmark in order to exclude the presence of ungapped quark 
matter in compact stars. The emissivity of a normal quark phase 
was first studied by Iwamoto \cite{Iwamoto:eb}, see also 
\cite{Iwamoto:1982,Burrows:1980ec,Burrows:1979uj}. Iwamoto showed
that the direct Urca process 
%$d\leftrightarrow u+e^-+\bar{\nu}$ 
is strongly suppressed in non-interacting quark matter, but that
Fermi liquid corrections in interacting matter lead to a fast
rate $\epsilon\sim \alpha_s T^6$, where $\alpha_s$ is the strong 
coupling constant. 

  It is also known, however, that unscreened color magnetic interactions 
lead to a breakdown of the Fermi liquid description at temperatures
$T_{nfl}\sim m \exp(-9 \pi/(4\alpha_s))$, where $m^2= N_F \alpha_s 
\mu^2/ \pi$ is the electric screening scale \cite{Holstein:1973,Baym:uj}. 
This scale is very small compared to ordinary QCD scales, but it 
is large compared to the temperature of neutron stars after the 
first minute or so. For $\mu=500$ MeV we have $T_{nfl}\sim 500$ keV. 
It was recently shown that non-Fermi liquid effects lead to large 
corrections of the specific heat of degenerate quark matter for
$T\ll T_{nfl}$ \cite{Ipp:2003cj}. In the present work we compute 
non-Fermi liquid corrections to the emissivity and study the 
cooling behavior of degenerate ungapped quark matter.

%%%%%%%%%%%%%%%%%%%%%%%%%%%%%%%%%%%%%%%%%%%%%%%%%%%%%%%%%%%%%%%%%%%%%%%%%
\section{Quark dispersion relation}
\label{sec_disp}
%%%%%%%%%%%%%%%%%%%%%%%%%%%%%%%%%%%%%%%%%%%%%%%%%%%%%%%%%%%%%%%%%%%%%%%%%

 In a Fermi gas of free quarks the direct Urca process is strongly 
suppressed. As a consequence, the rate is very sensitive to modifications 
of the quark dispersion relation. Iwamoto noticed that Fermi liquid 
corrections to the relation between the Fermi energy and the Fermi
momentum lead to a significant enhancement of the neutrino 
emissivity \cite{Iwamoto:eb}. Non Fermi liquid effects due to unscreened 
transverse gauge boson interactions dramatically alter the dispersion relation in the vicinity of the Fermi surface \cite{Baym:uj,Manuel:2000mk,Brown:2000eh,Boyanovsky:2000bc,Ipp:2003cj,Schafer:2003jn,Schafer:2004zf} and thereby could have a similar effect.
%cause the Fermi liquid description 
%to break down at temperatures $T\sim m\exp(-1/\gamma)$ where $\gamma=
%4\alpha_s/9$ 
In this section we wish to study the interplay of Fermi liquid and 
non-Fermi liquid effects in dense quark matter and derive the quark 
dispersion relation. 

%%%%%%%%%%%%%%%%%%%%%%%%%%%%%%%%%%%%%%%%%%%%%%%%%%%%%%%%%%%%%%%%%%%%%%%%
\begin{figure}
\begin{center}
\includegraphics[width=8cm]{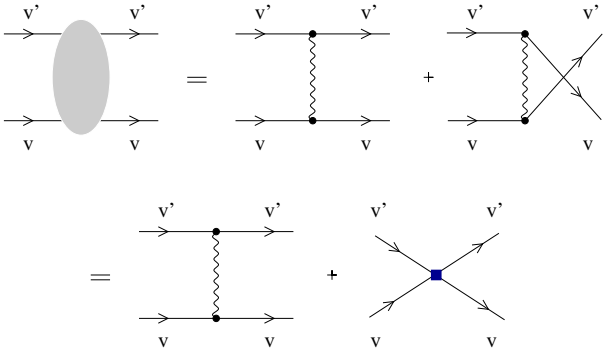}
\end{center}
\caption{Feynman diagrams that appear in the matching 
procedure for the forward scattering amplitude to 
leading order in the coupling constant. The upper 
panel shows the direct and exchange terms in QCD.
The lower panel shows the collinear and contact 
terms in the effective theory.}
\label{fig_forw}
\end{figure}
%%%%%%%%%%%%%%%%%%%%%%%%%%%%%%%%%%%%%%%%%%%%%%%%%%%%%%%%%%%%%%%%%%%%%%%%

 We are interested in the propagation of quarks in the 
vicinity of the Fermi surface. Since the momentum $p\sim 
v\mu$ is large, typical soft scatterings cannot change 
the momentum by very much and the velocity is approximately 
conserved. An effective field theory for particles and holes
moving with velocity $v=(1,\vec{v}_F)$ is given by 
\cite{Hong:2000ru,Schafer:2003jn}
\bea
\label{l_hdet}
{\cal L} &=& 
  \psi_{v}^\dagger \left( iv\!\cdot\!D+\delta\mu\right) \psi_{v} 
  - \frac{1}{4}G^a_{\mu\nu} G^a_{\mu\nu} + {\cal L}_{HDL} \nonumber \\
 && + \sum_{v',\Gamma} f_l^\Gamma R^l(v\cdot v') (\psi_v^\dagger\Gamma\psi_v)
       (\psi^\dagger_{v'}\Gamma\psi_{v'}) +\ldots .
\eea
Here, $D_\mu$ is the covariant derivative and $\delta\mu$ is a 
counterterm for the chemical potential. The effective theory contains 
two types of four fermion operators, 
corresponding to forward scattering $v+v'\to v+v'$ and the 
BCS process $v+(-v)\to v'+(-v')$ \cite{Schafer:2003jn}. Since 
we are interested in the normal phase only the forward scattering 
operators are included in equ.~(\ref{l_hdet}). The matrix $\Gamma$
determines the spin, color and flavor structure of the 
operator and $R^l(v\cdot v')$ is a set of orthogonal polynomials. 
Equ.~(\ref{l_hdet}) also contains the hard dense loop effective 
action
\be 
\label{S_hdl}
{\cal L}_{HDL} = -\frac{m^2}{2}\sum_v \,G^a_{\mu \alpha} 
  \frac{v^\alpha v^\beta}{(v\cdot D)^2} 
G^a_{\mu\beta},
\ee
which accounts for damping and screening caused by 
particle-hole pairs on the entire Fermi surface. 

 The parameters in the effective lagrangian are determined
by matching QCD Green functions near the Fermi surface. In 
order to determine the dispersion relation we have to match
forward scattering amplitudes, see Fig.~\ref{fig_forw}. To
leading order in the coupling constant the forward scattering
amplitude is the sum of a direct and an exchange term. In
the effective field theory the direct term is reproduced by 
the collinear interaction while the exchange term has to be
matched against a contact term \cite{Schafer:2003jn,Hands:2003rw}. 
The spin-color-flavor symmetric part is given by
\be 
\label{fl_c}
{\cal L} = f_0^s(\psi_v^\dagger \psi_v)
              (\psi^\dagger_{v'}\psi_{v'}), \hspace{1cm}
f_0^s= \frac{C_F}{4N_cN_f}\frac{g^2}{p_F^2},
\ee
with $C_F=(N_c^2-1)/(2N_c)$ and all other $f_i^s=0$. The parameters 
$v_F$ and $\delta\mu$ are determined by computing the contribution 
to the fermion dispersion relation from states far away from the Fermi 
surface. This can be done most easily using the hard dense loop 
approximation. We find 
\be 
 p_0 = |\vec{p}|+ \frac{m_f^2}{|\vec{p}|}, 
 \hspace{1cm} m_f^2=\frac{C_F \alpha_s}{2 \pi}\ \mu^2,
\ee
which gives
\be
\label{vf}
 v_F=1-\frac{C_F \alpha_s}{2 \pi}, \hspace{1cm}
 \delta\mu = \frac{C_F \alpha_s}{\pi}\mu.
\ee
These relations can also be derived using the Landau theory 
of Fermi liquids. Landau showed that Galilei invariance 
implies a relation between the effective interaction on 
the Fermi surface, encoded in the Landau parameters $f_l^s$,
and the parameters $v_F$ and $\delta\mu$. These arguments were 
generalized to the relativistic case by Baym and Chin 
\cite{Baym:1975va}. They show that
\be
\label{fl_rel}
v_F=\frac{p_F}{\mu}-\frac{Nf_1^s}{3}, \hspace{1cm}
p_F=\mu\left(1-\frac{Nf_0^s}{2}\right),
\ee
where $N=N_cN_f\mu^2/\pi^2$ is the density of states on the 
Fermi surface. Using equ.~(\ref{fl_c}) and (\ref{fl_rel}) gives 
equ.~(\ref{vf}). 

%%%%%%%%%%%%%%%%%%%%%%%%%%%%%%%%%%%%%%%%%%%%%%%%%%%%%%%%%%%%%%%%%%%%%%%%
\begin{figure}
\begin{center}
\includegraphics[width=8cm]{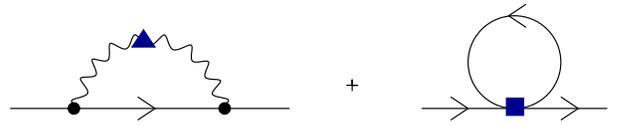}
\end{center}
\caption{Leading order contributions to the fermion self
energy in the effective theory. The triangle denotes a gluon
self energy insertion and the square is the four fermion
operator defined in Fig.~\ref{fig_forw}.}
\label{fig_sig}
\end{figure}
%%%%%%%%%%%%%%%%%%%%%%%%%%%%%%%%%%%%%%%%%%%%%%%%%%%%%%%%%%%%%%%%%%%%%%%%

 Finally, we can study loop corrections in the effective 
theory, see  Fig.~\ref{fig_sig}. The collinear loop gives
\be
\label{sig_1l}
\Sigma(\omega)=\frac{C_F \alpha_s}{3 \pi}\omega\log\left(
 \frac{\Lambda}{\omega}\right),
\ee
where $\omega=p_0-\mu$. We recently showed that this result does
not receive large logarithmic corrections of the form $g^{2n}\log^n
(\omega)$ as $\omega\to 0$ \cite{Schafer:2004zf}. The tadpole 
diagram is linearly divergent in the effective field theory. A 
naive estimate can be obtained by cutting the divergence off at 
$\Lambda\sim\mu$. This gives $\Sigma\sim f^0_s N\mu$ which 
agrees with the result for $\delta\mu$ given above. We can 
now summarize the results obtained in this section. The 
dispersion relation is 
\be
\label{disp-rel}
 \omega +\Sigma(\omega)=v_F l+ \delta\mu,
\ee	
where $l$ is the longitudinal momentum, $\Sigma(\omega)$ is 
given in equ.~(\ref{sig_1l}) and $v_F,\delta\mu$ given in 
equ.~(\ref{vf}).

%%%%%%%%%%%%%%%%%%%%%%%%%%%%%%%%%%%%%%%%%%%%%%%%%%%%%%%%%%%%%%%%%%%%%%%%%
\section{Neutrino emissivity}
\label{sec_neutrino}
%%%%%%%%%%%%%%%%%%%%%%%%%%%%%%%%%%%%%%%%%%%%%%%%%%%%%%%%%%%%%%%%%%%%%%%%%

%%%%%%%%%%%%%%%%%%%%%%%%%%%%%%%%%%%%%%%%%%%%%%%%%%%%%%%%%%%%%%%%%%%%%%%%
\begin{figure}
\begin{center}
\includegraphics[width=8cm]{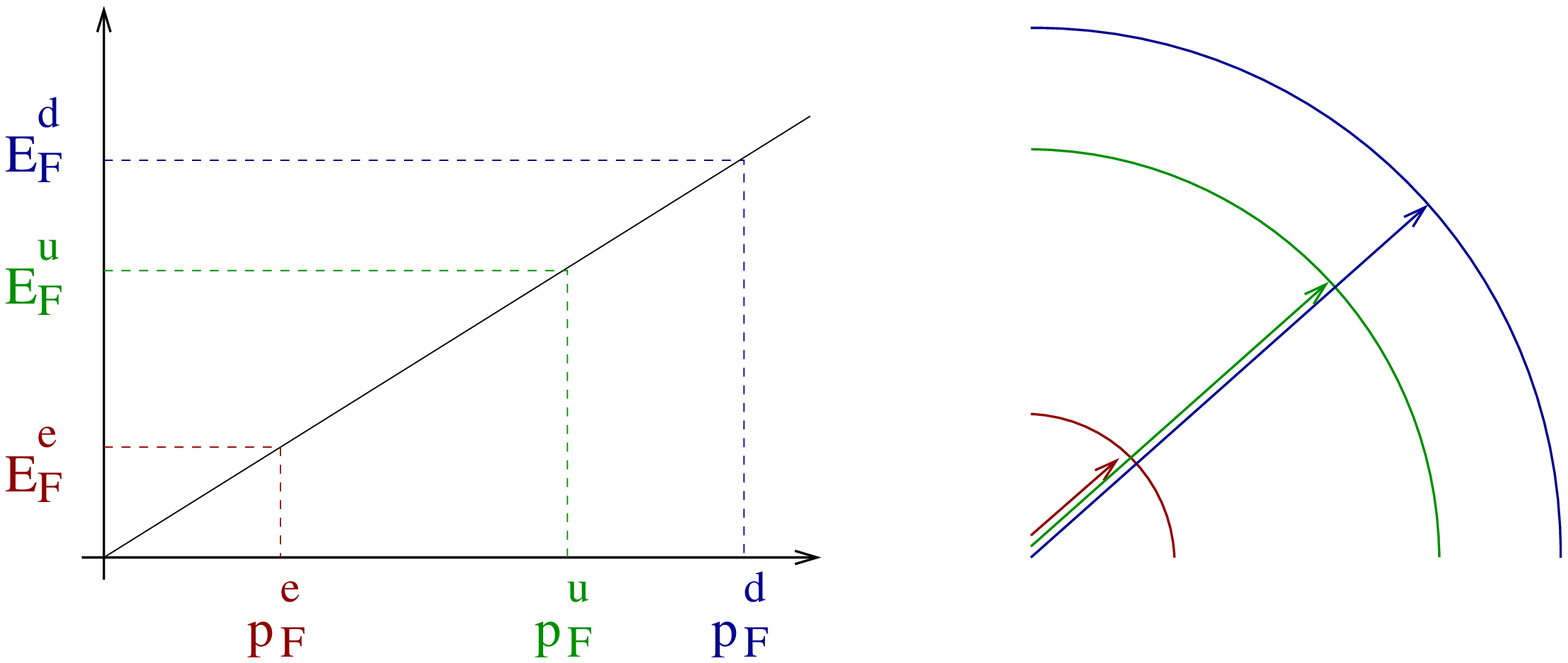}
\includegraphics[width=8cm]{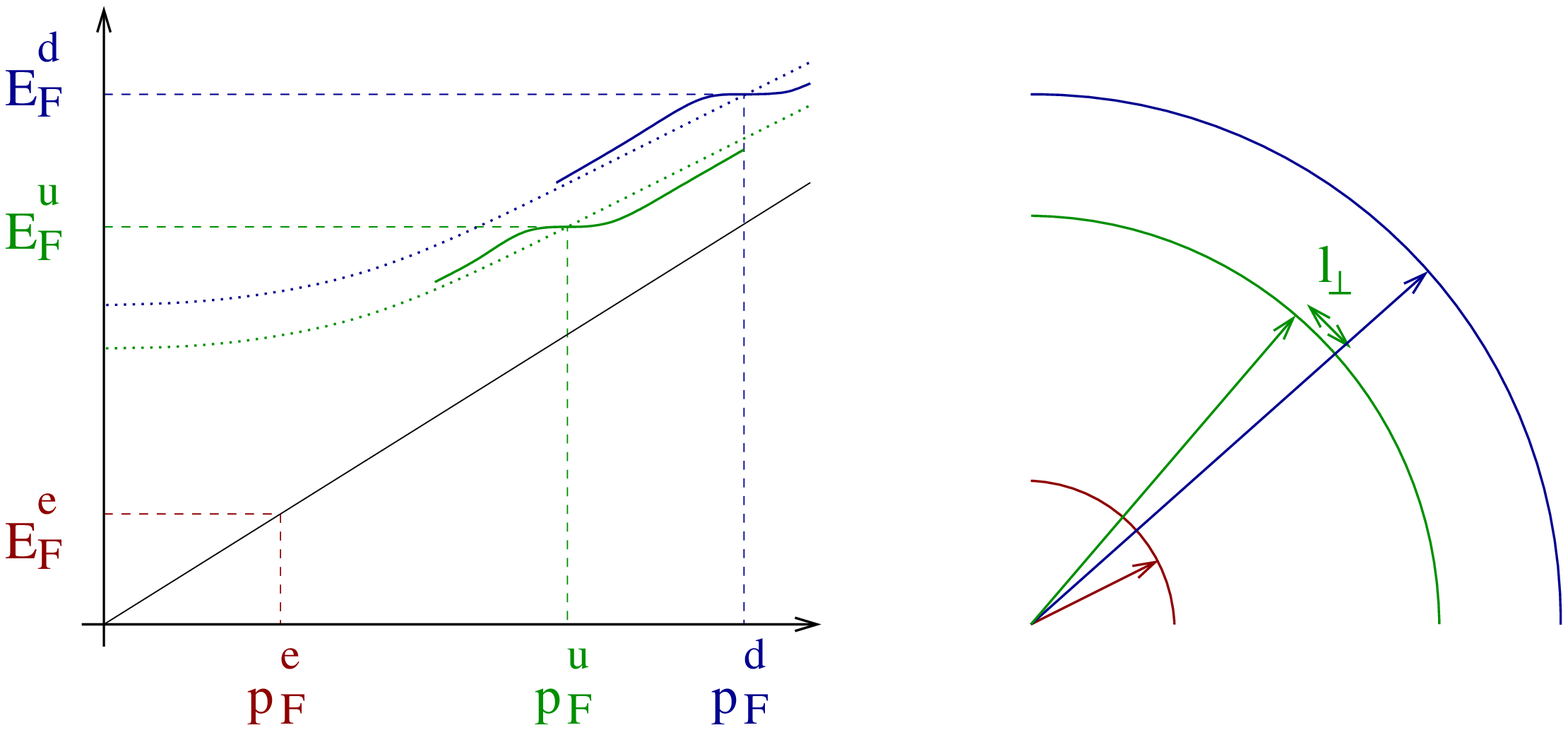}
\end{center}
\caption{Kinematics for the quark direct Urca process in the 
limit $T\sim E_\nu\ll\mu_e$. In a free quark gas (upper panel) energy momentum conservation  forces the quark and electron momenta to be collinear. If Fermi liquid corrections are taken into account (lower panel) the outgoing quark has a
non-zero transverse momentum $\l_\perp^2\sim \alpha_s\mu_e^2$. The dashed dispersion relations give the HDL result whereas the solid lines show the change when non-Fermi liquid corrections are included. These lead to a flattening of the dispersion relation in the vicinity of the Fermi surface.}
\label{fig_kin}
\end{figure}
%%%%%%%%%%%%%%%%%%%%%%%%%%%%%%%%%%%%%%%%%%%%%%%%%%%%%%%%%%%%%%%%%%%%%%%%

 The dominant contribution to the emission of neutrinos is given 
by the quark analogs of $\beta$-decay ($\beta$) and electron 
capture ($ec$)
\bea
  \label{beta-decay}
  d &\rightarrow& u + e^- + \bar\nu_e, \\
  \label{electron-capture}
  u + e^- &\rightarrow& d + \nu_e.
\eea
It is straightforward to introduce weak interactions into the
effective theory. The charged current interaction is given 
by 
\be
{\cal L} = \frac{g_2}{\sqrt{2}}\cos\theta_c\psi^\dagger \tau^\pm
 v\cdot W^\mp \psi
\ee
where $\cos\theta_c$ is the Cabbibo angle and $g_2$ is related 
to the Fermi coupling by $G_F/\sqrt{2}=g_2^2/(8M_W^2)$. The dependence 
on the Cabbibo angle suppresses the processes involving the  strange 
quark \cite{Iwamoto:eb}. Therefore we will neglect the neutrino emission of strange quarks. We have seen in the previous section that 
almost collinear gluon exchanges can generate large logarithmic
corrections to the fermion self energy. This raises the question whether
gluon corrections to the weak interaction vertex have to be taken 
into account. The problem of vertex corrections was studied in 
\cite{Brown:2000eh,Schafer:2003jn}. It was shown that the vertex
receives large logarithmic corrections in the time-like regime, 
but not in the space-like regime. The coefficient of the logarithm 
in the time-like regime is exactly equal to the logarithmic term 
in the fermion self energy. In the fermion self energy problem
the vertex is space-like and does not have to be renormalized. 
However, since the wave function renormalization is logarithmically
divergent the effective coupling constant goes to zero near the 
Fermi surface. In the neutrino emission problem the vertex is 
time-like and receives logarithmic corrections. However, since 
the vertex correction is equal to the field renormalization the 
effective coupling constant does not change. 

 The neutrino emissivity is given by the total energy loss due to 
neutrino emission averaged over the initial quark helicities  and 
summed over the final state phase space and helicities
\begin{widetext}
\bea
  \epsilon &\equiv&   N_c  \sum_{\sigma_u, 
 \sigma_d, \sigma_ e} 
 \int \frac{d^3 p_d}{(2 \pi)^3} \frac{1}{2 E_d} 
 \int \frac{d^3 p_u}{(2 \pi)^3} \frac{1}{2 E_u} 
 \int \frac{d^3 p_e}{(2 \pi)^3} \frac{1}{2 E_e} 
 \int \frac{d^3 p_\nu}{(2 \pi)^3} \frac{1}{2 E_\nu} \, E_\nu \\ \nonumber 
  && \qquad \qquad \qquad \qquad \cdot 
   \left( |M_{\beta}|^2 (2 \pi)^4 \delta^{(4)}(p_d-p_u-p_e-p_\nu) 
   n(p_d) (1-n(p_u)) (1-n(p_e)) \right. \\ \nonumber
  && \left. \qquad \qquad \qquad \qquad \qquad \quad 
  + |M_{ec}|^2 (2 \pi)^4 \delta^{(4)} (p_u+p_e-p_d-p_\nu) 
   n(p_u) n(p_e) (1-n(p_d)) \right)
\eea
\end{widetext}
The weak matrix element for the $\beta$ and $ec$ processes is 
given by 
\be
\tfrac{1}{2} \sum_{\sigma_u, \sigma_d, \sigma_ e} 
  | M_{\beta / ec}|^2 = 64G_F^2\cos^2\theta_c p_F^2 
                              (v\cdot p_e)(v\cdot p_\nu),
\ee
where $p_e,p_\nu$ are the momenta of the electron and the neutrino. 
Weak processes establish $\beta$ equilibrium in the star. In three 
flavor quark matter with a massive strange quark the resulting electron 
chemical potential is small. In the following we 
shall assume that $(T\sim E_\nu) \ll (\mu_e \sim E_e) \ll p_F$. This 
assumption is appropriate in all cases except during the first few 
seconds of the proto-neutron star evolution.

 In this case we can neglect the neutrino momentum when 
applying the energy-momentum conservation relation to the matrix 
element. As a consequence we find $(v\cdot p_\nu)\simeq E_\nu$ after 
averaging over the direction of the outgoing neutrino. The matrix 
element is mainly determined by the factor $(v\cdot p_e)$. To leading 
order in the effective theory the weak decay is exactly collinear 
and $(v\cdot p_e)=(E_e-v_Fl_e)=0$ up to terms of order $O(T/\mu_e)$,
see Fig.~\ref{fig_kin}. If corrections to the dispersion relation 
are taken into account we get 
\be
 (v\cdot p_e)\simeq(1-v_F)l_e +\frac{l_\perp^2}{2l_e}
  \simeq \delta\mu_d-\delta\mu_u
  \simeq \frac{C_F\alpha_s}{\pi} \mu_e ,
\ee
where $l_\perp$ is the transverse momentum of the electron. 
This result agrees with Iwamoto's result. Non-Fermi liquid
corrections only appear in the phase space integral. To leading order 
in the $T/\mu$ the sum of the rates for electron capture and $\beta$
decay is given by
\begin{widetext}
\begin{align}
\label{emi}
\epsilon = &\frac{6 G_F^2 \cos^2 \theta_c}{\pi^5} T^6 
  \int_{-\infty}^\infty d x_d \int_{-\infty}^\infty d x_u 
  \int_0^\infty d x_\nu \, x_\nu^3 \, n(x_d) n(-x_u) 
  n(x_u\!-\!x_d\!+\!x_\nu)  
\nonumber \\ &\qquad \qquad \qquad \qquad \qquad \cdot 
\left[  
  \frac{\partial p(E_d)}{\partial E_d}  
  \frac{\partial p(E_u)}{\partial E_u} 
  \frac{C_F\alpha_s}{\pi} \mu_e E_u E_d 
\right]_{E_i \rightarrow \mu_i+T x_i} .
\end{align}
\end{widetext}
%\begin{widetext}
%\begin{align}
%\label{emi}
%\epsilon = &\frac{3 G_F^2 \cos^2 \theta_c}{2 \pi^5} T^6 \int_{-\infty}^\infty d x_d %\int_{-\infty}^\infty d x_u \int_0^\infty d x_\nu \, x_\nu^3 \, n(x_d) n(-x_u) n(x_u\!-\!%x_d\!+\!x_\nu)  \nonumber \\ 
%&\qquad \qquad \qquad \qquad \cdot \left[ \frac{p(E_d)}{E_d} \frac{\partial %p(E_d)}{\partial E_d} \frac{p(E_u)}{E_u} \frac{\partial p(E_u)}{\partial E_u} \left( %p(E_u)^2-E_u^2-p(E_d)^2+E_d^2 \right) ( E_d+E_u ) \right]_{E_i \rightarrow %\mu_i+T x_i}
%\end{align}
%\end{widetext}
The expression in the square brackets is determined by the quark 
dispersion relation given in equ.~(\ref{disp-rel}). Terms of 
$O(\alpha_s(\alpha_s\log(T))^n)$ with $n=0,1,2$ are independent 
of $\log(x_i)$ and involve the integral 
\bea
 && \int_{- \infty}^\infty d x_d \int_{- \infty}^\infty  d x_u 
 \int_{0}^\infty d x_\nu x_\nu^3 \, n(x_d) n(-x_u) \nonumber  \\
  && \quad\quad\quad\quad \mbox{}\cdot   n(x_u-x_d+x_\nu) 
   = \frac{457 \pi^6}{5040} .
\eea
At leading order in  $T/\mu$ the neutrino emissivity from 
the quark direct Urca process is given by
\begin{equation}
\label{emissivity}
\epsilon \approx \frac{457}{630} \, G_F^2 \, 
     \cos^2 \theta_c \, \alpha_s \, \mu_q^2 \, \mu_e \, T^6 
     \left( 1\!+\!\frac{C_F \,  \alpha_s}{3 \pi} 
     \log \left( \frac{\Lambda}{T} \right) \right)^2 .
\end{equation}
The first term is the standard result by Iwamoto \cite{Iwamoto:eb}, 
and the logarithmic terms are non-Fermi liquid corrections. 
We note that these terms have to be included because at very low 
temperature $\alpha_s\log(T)$ becomes large compared to 
one. We also note that if the scale inside the logarithm is on
the order of the screening scale, $\Lambda\sim g\mu$, then 
$\alpha_s(\mu)\log(\Lambda/T)$ stays finite in the limit 
$\mu\to\infty$ at fixed $T$.

%%%%%%%%%%%%%%%%%%%%%%%%%%%%%%%%%%%%%%%%%%%%%%%%%%%%%%%%%%%%%%%%%%%%%%%%%
\begin{figure*}
\includegraphics[width=12cm]{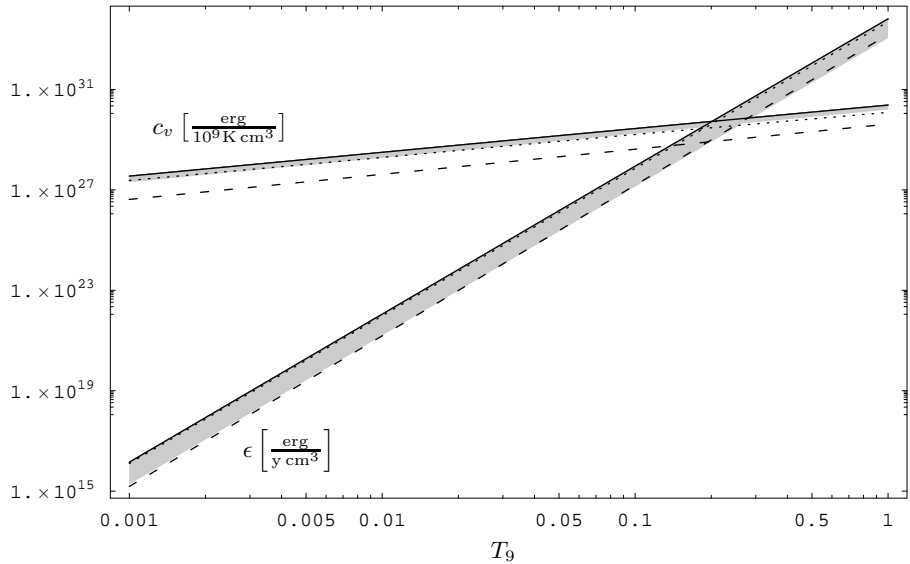}
\flushleft \vspace*{-0.5cm} \hspace*{9.3cm}  $T_9$
\flushleft \vspace*{-6.3cm} \hspace*{4.8cm}  $c_v \left[ 
   \frac{\mathrm{erg}}{\mathrm{10^9 K} \, \mathrm{cm}^3} \right]$
   \flushleft \vspace*{3.5cm} \hspace*{6cm} $\epsilon 
   \left[ \frac{\mathrm{erg}}{\mathrm{y} \, \mathrm{cm}^3} \right]$
\vspace*{1,3cm}
\caption{Neutrino emissivity $\epsilon$ and specific heat $c_v$ of 
quark matter. The dashed line shows the Fermi liquid results and
the dotted lines show the anomalous corrections. The solid line 
gives the sum of the two contributions and the gray band shows
an estimate of the uncertainties.}
\label{sh_ne}
\end{figure*}
%%%%%%%%%%%%%%%%%%%%%%%%%%%%%%%%%%%%%%%%%%%%%%%%%%%%%%%%%%%%%%%%%%%%%%%%%

%%%%%%%%%%%%%%%%%%%%%%%%%%%%%%%%%%%%%%%%%%%%%%%%%%%%%%%%%%%%%%%%%%%%%%%%%
\section{Compact star cooling}
\label{sec_cooling}
%%%%%%%%%%%%%%%%%%%%%%%%%%%%%%%%%%%%%%%%%%%%%%%%%%%%%%%%%%%%%%%%%%%%%%%%%

 In this section we wish to study the impact of non-Fermi liquid effects
on the cooling history of an isolated quark phase. Our aim is not to 
provide a thorough analysis of the cooling behavior of an actual quark 
or hybrid star, but to give a numerical estimate of the size of 
the non-Fermi liquid corrections. The thermal evolution of the star
is governed by the neutrino emissivity, the specific heat and 
the thermal conductivity. Non-Fermi liquid corrections to the specific
heat were initially considered by Holstein et al.~\cite{Holstein:1973} 
in the case of QED. The calculation was recently refined and extended
to QCD by Ipp et al.~\cite{Ipp:2003cj}. They find 
\be
\label{spec-heat}
  c_v = \frac{N_c N_f}{3} \mu_q^2 \, T 
  \left( 1 + \frac{C_F \, \alpha_s}{3 \pi} 
  \log \left( \frac{\Lambda}{T} \right) \right), 
\ee
where the first term is the free gas result and the second term is the 
non-Fermi liquid correction. Ipp et al.~also determined the scale inside 
the logarithm as well as fractional powers of $T$. From the complete
$O(\alpha_s)$ result we find $\Lambda \simeq 0.28 \, m$ where $m^2=
N_f \alpha_s \mu^2/\pi$ is the dynamical mass. 

 The thermal conductivity of degenerate quark matter was studied 
by Pethick and Heiselberg \cite{Heiselberg:1993cr}. Their result
suggests that equilibration is fast and that the quark phase is 
isothermal. In this case the cooling behavior is governed by 
\be
  \frac{\partial u}{\partial t} = \frac{\partial u}{\partial T} 
  \frac{\partial T}{\partial t} = c_v(T) \frac{\partial T}{\partial t} 
    = - \epsilon(T),
\label{eq_cool}
\ee
where $u$ is the internal energy, $t$ is time and we have assumed 
that there is no surface emission. Without non-Fermi liquid effects
we have $\epsilon\sim T^6$ and $c_v\sim T$. In this case the 
temperature scales as $T \propto 1/t^\frac{1}{4}$. With logarithmic
corrections included there is no simple analytic solution and 
we have studied equ.~(\ref{eq_cool}) numerically. 

 We take the quark chemical potential to be $\mu_q\!=\!500$ MeV
corresponding to densities $\rho_B \approx 6\rho_0$ where $\rho_0$
is nuclear matter saturation density. We note that both $c_v$
and $\epsilon$ are proportional to $\mu^2$ and the main 
dependence of the cooling behavior on $\mu$ cancels. We 
evaluate the strong coupling constant using the one loop 
renormalization group solution at a scale $\mu$. We take 
the scale parameter to be $\Lambda_{QCD}\!=\!250$ MeV which 
gives $\alpha_s\simeq 1$ at $\mu=500$ MeV. It is clear that 
the naive use of perturbation theory is in doubt if the coupling 
is this large. In practice we estimate the uncertainty by 
varying $\alpha_s$ between 1 and 0.4 which is the value used 
by Iwamoto \cite{Iwamoto:eb}. We take the scale of non-Fermi
liquid effects to be $\Lambda=0.28m$ as explained above and
assess the uncertainty by varying $\Lambda$ within a factor 
of 2. Finally, we took the initial temperature to be $T_0 \!=\!15$ MeV.

  The electron chemical potential is determined by the requirements 
of charge neutrality and $\beta$-equilibrium. In a non-interacting 
quark gas we find $\mu_e\simeq m_s^2/(4p_F)$. With a strange quark 
mass $m_s\!=\!150$ MeV this relation gives $\mu_e\! \approx\! 11$ MeV. This
result, however, is very sensitive to interactions. To first order
in $\alpha_s$ the chemical potential for a massive strange quark is 
\cite{Freedman:1977gz,Duncan:1983wy}
\begin{equation}
\label{mus}
  \mu_s=E_{Fs}^0+\!\frac{2 \alpha_s}{3 \pi} 
    \left( p_{Fs}-\frac{3 m_s^2}{E_{Fs}^0} 
   \log \left( \frac{p_{Fs}+E_{Fs}^0}{m_s} \right) \right) 
\end{equation}
where $E_{Fs}^0=\sqrt{p_{Fs}^2\!+\!m_s^2}$. The important point is
that the $O(\alpha_sm_s^2)$ term is negative and enhanced by a 
large logarithm $\log(p_F/m_s)$. The sign is related to the fact 
that the correlation energy changes sign in going from the 
relativistic to the non-relativistic limit. 

 Equation (\ref{mus}) implies that the strange quark chemical 
potential can become equal to or even smaller than the up quark
chemical potential. To leading order in $m_s^2/p_F^2$ the electron
chemical potential is given by 
\be
\label{mue}
\mu_e \simeq \frac{m_s^2}{4p_F}\left( 
 1 -\frac{4\alpha_s}{\pi} \log\left(\frac{2p_F}{m_s}\right)\right).
\ee
For the values of the parameters given above this equation gives
a negative electron chemical potential $\mu_e\!\approx\!-16$ MeV. In 
this case the quark phase contains a Fermi sea of positrons and the 
quark direct Urca process is 
\begin{equation}
\label{revurca}
u \rightarrow d + e^+ +\nu \qquad , \qquad 
d + e^+ \rightarrow u + \bar\nu \, .
\end{equation}
The neutrino emissivity is again governed by equ.~(\ref{emissivity})
where $\mu_e$ has to be replaced by $-\mu_e$. We observe that despite 
the large correction to $\mu_e$ the emissivity is not strongly 
affected. The large variation in $\mu_e$ when perturbative 
corrections are included implies, however, that the electron
chemical potential is very uncertain. In particular, there is 
a possibility that $\mu_e$ is much smaller than $m_s^2/(4p_F)$. 
If $\alpha_s\mu_e<T$ then the neutrino emissivity is no longer 
proportional to $\alpha_s\mu_e T^6$ but to $T^7$ \cite{Burrows:1980ec}. 
 
%%%%%%%%%%%%%%%%%%%%%%%%%%%%%%%%%%%%%%%%%%%%%%%%%%%%%%%%%%%%%%%%%%%%%%%%%
\begin{figure*}
\includegraphics[width=12cm]{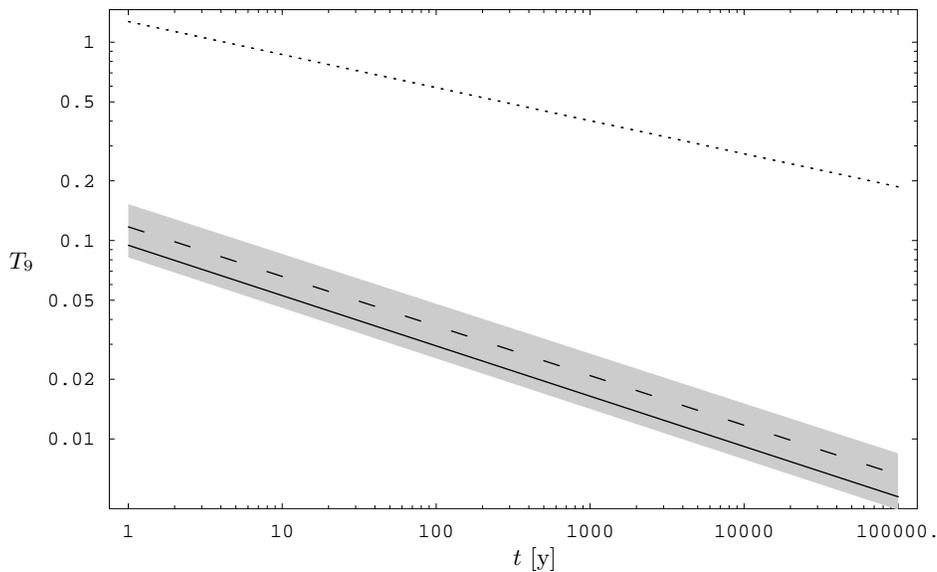}
\flushleft \vspace*{-0.5cm} \hspace*{9.1cm}  $t$ [y]
\flushleft \vspace*{-4.6cm} \hspace*{2.4cm} $T_9$\vspace*{3.8cm}
\caption{Cooling behavior of ungapped quark matter. We show 
the temperature $T_9$ in units of $10^9$ K as a function of the age 
of the star in years. The dashed line shows the Fermi liquid 
result whereas the solid line gives the result including non-Fermi 
liquid effects with the estimated uncertainty range. Although the 
non-Fermi liquid corrections to both the specific heat and the neutrino 
emissivity are significant, there is only a modest reduction in the 
temperature at late times. However, for both cases the cooling of 
quark matter is considerably faster than the cooling of neutron matter 
via the modified Urca process given by the dotted line.}
\label{cooling}
\end{figure*}
%%%%%%%%%%%%%%%%%%%%%%%%%%%%%%%%%%%%%%%%%%%%%%%%%%%%%%%%%%%%%%%%%%%%%%%%%

 In the following we shall use the value $\mu_e\!=\!16$ MeV corresponding 
to interacting quarks. In Fig.~\ref{sh_ne} we show the temperature 
dependence of both the neutrino emissivity and the specific heat 
(solid) compared to the Fermi liquid result (dashed). The gray band 
shows an estimate of the uncertainties which are dominated by the 
uncertainty in the value of the strong coupling. For both $c_v$ and 
$\epsilon$ the anomalous logarithmic terms (dotted) dominate in the 
relevant temperature range and exceed the Fermi liquid result considerably. 
The cooling behavior is controlled by the ratio $\epsilon/c_v$.
Since $\epsilon\sim\log^2(T)$ and $c_v\sim\log(T)$ this ratio is
logarithmically enhanced. However, because the temperature at late
times scales roughly as the fourth root of the numerical coefficient
in $\epsilon/c_v$ this logarithmic enhancement only translates into 
a modest reduction of the temperature. This can be seen in more 
detail in Fig.~\ref{cooling}. We observe that compared to the Fermi 
liquid result (dashed) the non-Fermi liquid effects (solid) lead 
to a reduction of the temperature at late times which is nearly 
independent of time. The magnitude of the effect is on the order 
of $20 \%$. For comparison, we also show the cooling behavior of
normal nuclear matter via the modified Urca process $n + n \rightarrow 
n + p + e^- + \bar \nu$ \cite{Friman:zq}. We have chosen the same
density and initial temperature and the effective baryon masses given in \cite{Page:2004fy}. We clearly see the difference 
between the fast $\sim T^6$ quark direct Urca process and the 
slow $\sim T^8$ modified Urca process. 

%%%%%%%%%%%%%%%%%%%%%%%%%%%%%%%%%%%%%%%%%%%%%%%%%%%%%%%%%%%%%%%%%%%%%%%%%
\section{Summary and Discussion}
\label{sec_sum}
%%%%%%%%%%%%%%%%%%%%%%%%%%%%%%%%%%%%%%%%%%%%%%%%%%%%%%%%%%%%%%%%%%%%%%%%%

In this work, we have discussed the influence of non-Fermi liquid 
effects on the cooling behavior of compact stars containing quark
matter in the normal phase. Non-Fermi liquid effects lead to a 
logarithmic enhancement in both the neutrino emissivity and the 
specific heat. The net result of these two effects is a mild logarithmic
enhancement in the cooling rate. As our rate is even larger than 
the Iwamoto rate we confirm and sharpen earlier bounds on the 
existence of ungapped quark matter in neutron stars 
\cite{Blaschke:2000dy,Page:2000wt,Prakash:2000jr,Slane:2002ta}. 
More quantitative statements will require detailed studies of realistic 
models in which the quark core is in contact with a hadronic phase or 
an atmosphere. This is beyond the scope of our investigation. 

%%%%%%%%%%%%%%%%%%%%%%%%%%%%%%%%%%%%%%%%%%%%%%%%%%%%%%%%%%%%%%%%%%%%%%%%%
\begin{table}
\begin{center}
\begin{tabular}{|c|c|c|} \hline
phase &  emissivity  & specific heat \\ \hline\hline  
normal phase &  $\epsilon\sim \alpha_s^3 T^6\log^2(T)$  & 
   $c_v \sim \alpha_s \mu^2 T\log(T)$  \\ \hline 
CFL   &  $\epsilon \sim T^{3/2} \exp(-m_K/T)$    &
   $c_v \sim T^3$  \\ \hline
2SC   &   $\epsilon\sim \alpha_s T^6$   &
    $c_v \sim \mu^2 T$  \\ \hline 
2SC+1SC   &   $\epsilon\sim \alpha_s T^6$   &
    $c_v \sim \mu^2 T$  \\ \hline 
$[1{\rm SC}]^3$   &  $\epsilon\sim \alpha_s T^6$   &
    $c_v \sim \mu^2 T$  \\ \hline 
gCFL  & $\epsilon\sim \alpha_s T^{5.5}$  & 
    $c_v \sim \mu^2 \sqrt{\Delta T}$  \\ \hline 
gCFLK & $\epsilon \sim  \alpha_s T^8$ &
    $c_v \sim \mu^2 T$   \\ \hline 
g2SC  &  $\epsilon\sim \alpha_s T^6$   & 
    $c_v \sim \mu^2 T$  \\ \hline
\end{tabular}
\vspace{0.2cm}
\caption{Dominant cooling rates in different phases of quark matter. 
This table is a summary of the discussion in Sect.~\ref{sec_sum}.
In the entry $[1{\rm SC}]^3$ we have assumed that ungapped modes 
exist and the Urca process is allowed. This is not true in the CSL 
phase. In the gCFL phase we have used the estimate of Kouvaris 
et al.~and in the gCFLK phase we have assumed that only slow processes
such as neutrino bremsstrahlung are allowed. In the g2SC phase we have 
assumed that the Fermi liquid direct Urca rate can be used. In 
both the gCFL and g2SC phase we have ignored possible instabilities
signaled by a negative current-current correlation function. }
\label{tab_cool}
\end{center}
\end{table}
%%%%%%%%%%%%%%%%%%%%%%%%%%%%%%%%%%%%%%%%%%%%%%%%%%%%%%%%%%%%%%%%%%%%%%%%%

 Let us now discuss the importance of non-Fermi liquid effects in 
partially gapped color superconducting phases, see Table \ref{tab_cool}. 
The simplest case is the 2SC phase. The 2SC phase can arise when the 
difference between the Fermi momenta of the strange quark and the up 
and down quarks is too large for strange-non-strange pairing to occur
\cite{Schafer:1999pb,Alford:1999pa}. There also regions in the 
phase diagram where $us$ (dSC) or $ds$ (uSC) pairing might occur 
\cite{Iida:2003cc,Fukushima:2004zq,Ruster:2004eg}. The 
2SC phase is characterized by a partial Higgs mechanism. Color 
$SU(3)$ is broken according to $SU(3)\to SU(2)$ and five out of 
eight gluons acquire a mass. The gapless fermions of the third 
color interact via screened gauge bosons. As a consequence, there 
are no non-Fermi liquid corrections in either the specific heat 
or the neutrino emissivity and the standard result of Iwamoto
applies \cite{Iwamoto:eb}. 

 If the 2SC phase occurs in quark matter with three flavors then 
the unpaired quark flavor can form a spin one condensate. Spin
one condensates are interesting from the point of view of cooling
because the typical gaps are much smaller than the spin zero gap. 
In particular, we can have $T_c \simeq T_{nfl}$ and non-Fermi liquid
effects in the specific heat are important. Whether or not 
gapless modes occur below $T_c$ depends on the exact nature 
of the spin one condensate \cite{Schafer:2000tw,Schmitt:2003aa}.
In a phase with CSL pairing in both $LL$ and $LR$ chirality 
channels the order parameter is isotropic and no gapless modes 
exist. In all other phases not all fermions are gapped and the 
order parameter has nodes on Fermi surface. If spin one pairing 
takes place in all flavor sectors (referred to as $[1{\rm SC}]^3$
in Table \ref{tab_cool}) these modes will dominate the cooling
behavior.

 The cooling behavior of the recently proposed gapless CFL 
\cite{Alford:2003fq} and 2SC \cite{Shovkovy:2003uu} phases is a 
more difficult question. In the gCFL phase there are gapless charged 
and neutral modes and the direct Urca process is possible. Alford 
et al.~\cite{Alford:2004zr} have argued that the (almost) quadratic 
dispersion relation of one of the gapless modes leads to an enhancement 
of the cooling rate by a factor $\sqrt{T}$. A similar enhancement occurs 
in the specific heat \cite{Alford:2003fq}. Kryjevski and Sch{\"a}fer 
studied quark modes in a kaon condensed CFL phase \cite{Kryjevski:2004jw}.
They find at most one gapless mode. As a consequence, the dominant 
cooling mechanism is expected to be a slow process such as neutrino 
bremsstrahlung. However, there are several light modes that will
contribute to neutrino transport at MeV temperatures. In the g2SC
phase there are gapless up and down quark excitations and the 
direct Urca process is possible. 

 We should stress that our discussion of neutrino emission 
from the g2SC and gCFL phase ignores potential instabilities
related to a negative current-current correlation function 
\cite{Wu:2003,Huang:2004bg,Casalbuoni:2004tb}. Wu and Yip
as well as Giannakis and Ren \cite{Wu:2003,Giannakis:2004pf}
have argued that these instabilities indicate that the correct 
ground state is an inhomogeneous superconductor of the type 
discussed by Larkin, Ovchninkov, Fulde and Ferrell 
\cite{Larkin:1964,Fulde:1964,Alford:2001ze}. The LOFF 
phase also contains gapless fermions and the calculation
of the neutrino emissivity from this phase in an 
important problem for future studies.

{\bf Note added in proof:} \\
Since the first version of this work the scale in the logarithm of equ. (\ref{sig_1l}) has been computed by Gerhold and Rebhan \cite{Gerhold:2005uu} and the result $\Lambda_\Sigma \approx 8 m$ turned out to be substantially larger than the corresponding scale in the specific heat $\Lambda_c \approx 0.28 m$ we have employed as an estimate for the scale in the emissivity. This leads quantitatively to a further enhancement of the neutrino emissivity but does not change the cooling behavior significantly.

\acknowledgments
We would like to thank C.~Kouvaris, M.~Prakash, K.~Rajagopal, 
and S.~Reddy for useful discussions and Q. Wang for pointing out an error
in equ. (12) of the first version of this manuscript. This work was begun at the 
Institute for Nuclear Theory during the workshop on ``QCD 
and Dense Matter: From Lattices to Stars''. We thank the INT for 
hospitality. This work was supported in part by US DOE grant 
DE-FG-88ER40388.

%%%%%%%%%%%%%%%%%%%%%%%%%%%%%%%%%%%%%%%%%%%%%%%%%%%%%%%%%%%%%%%%%%%%%%%%%

\end{document}